\documentclass[aps,10pt,twocolumn,tightenlines,superscriptaddress,floatfix]{revtex4-1}  
\usepackage{graphicx}
\usepackage[T1]{fontenc}
\usepackage[utf8]{inputenc} 
\usepackage{times} 
\usepackage{dcolumn}

\usepackage{natbib}

\begin{document}

\title{Electronic band gaps of confined linear carbon chains ranging from polyyne to carbyne\\}

\author{Lei Shi}
 \affiliation{%
 University of Vienna, Faculty of Physics, 1090 Wien, Austria\\}%

\author{Philip Rohringer}%
\affiliation{%
 University of Vienna, Faculty of Physics, 1090 Wien, Austria\\}%
\affiliation{%
 University of Antwerp, Experimental Condensed Matter Physics Laboratory, B-2610 Antwerp, Belgium \\}%

\author{Marius Wanko}
\affiliation{
Nano-Bio Spectroscopy Group and ETSF, Dpto. Material Physics, Universidad del Pa\'{i}s Vasco, 20018 San Sebasti\'{a}n, Spain\\ }%

\author{Angel Rubio}
\affiliation{
Nano-Bio Spectroscopy Group and ETSF, Dpto. Material Physics, Universidad del Pa\'{i}s Vasco, 20018 San Sebasti\'{a}n, Spain\\ }%

\affiliation{
Max Planck Institute for the Structure and Dynamics of Matter, Hamburg, Germany\\ }%

\author{S{\"o}ren Wa{\ss}erroth}
\affiliation{
 Freie Universit{\"a}t Berlin, Department of Physics, Arnimallee 14, 14195 Berlin, Germany\\ }%

\author{Stephanie Reich}
\affiliation{
 Freie Universit{\"a}t Berlin, Department of Physics, Arnimallee 14, 14195 Berlin, Germany\\ }%

\author{Sofie Cambr{\'e}}%
 \affiliation{%
 University of Antwerp, Experimental Condensed Matter Physics Laboratory, B-2610 Antwerp, Belgium \\}%

\author{Wim Wenseleers}%
 \affiliation{%
 University of Antwerp, Experimental Condensed Matter Physics Laboratory, B-2610 Antwerp, Belgium \\}%

\author{Paola Ayala}
 \affiliation{%
 University of Vienna, Faculty of Physics, 1090 Wien, Austria\\}%
 
 \author{Thomas Pichler} \email{thomas.pichler@univie.ac.at}
 \affiliation{%
 University of Vienna, Faculty of Physics, 1090 Wien, Austria\\}%

\begin{abstract}
Ultra long linear carbon chains of more than 6000 carbon atoms have recently been synthesized within double-walled carbon nanotubes, and they show a promising new route to one--atom--wide semiconductors with a direct band gap. Theoretical studies predicted that this band gap can be tuned by the length of the chains, the end groups, and their interactions with the environment. However, different density functionals lead to very different values of the band gap of infinitely long carbyne. In this work, we applied resonant Raman excitation spectroscopy with more than 50 laser wavelengths to determine for the first time the band gap of long carbon chains encapsulated inside DWCNTs. The experimentally determined band gaps ranging from 2.253 to 1.848 eV follow a linear relation with Raman frequency. This lower bound is the smallest band gap of linear carbon chains observed so far. The comparison with experimental data obtained for short chains in gas phase or in solution demonstrates the effect of the DWCNT encapsulation, leading to an essential downshift of the band gap. This is explained by the interaction between the carbon chain and the host tube, which greatly modifies the chain's bond length alternation.
\end{abstract}

\maketitle
\section{\label{sec:level1}Introduction}
One-dimensional linear carbon chains (LCCs) possess unique properties, one of which is a direct band gap that is tuneable by the length of the chains \citep{Kertesz78JCP,Casari15N,Yang06JPCA}. Despite their apparently simple structure, the properties of long LCCs (LLCCs) are hard to calculate because long-range electron exchange and correlation effects lead to a Peierls distortion, yielding a structure with significant bond-length-alternation (BLA) (which moreover depends strongly on end-capping effects). Their experimental observation has long remained elusive because of the high reactivity of the chains. Until recently, only relatively short chains (polyynes up to 44 carbon atoms) could be synthesized and stabilized \citep{Chalifoux10NC}, showing a linear relation between band gap and inverse chain length \citep{Agarwal13JRS}. Being only one atom wide, the electronic spectrum of these short LCCs furthermore depends drastically on the local environment and on the different end groups that are implemented to stabilize the chains, thus providing a whole range of tools to continuously tune the band gap of these materials in a very wide range through length, environment, and end groups \citep{Agarwal13JRS}. Very recently, we succeeded in stabilizing ultra long LCCs of more than 6000 carbon atoms within double-walled carbon nanotubes (DWCNTs) \citep{Shi16NM}, which now allows us to determine also the electronic band gap of such ultra long chains.

Tuning of the band gap by changing the material's properties plays a crucial role in the design of new semiconductor devices. In the past, fundamental research has focused mainly on other (quasi) 1D systems, in particular on carbon nanotubes, which show a band gap that is strongly dependent on their chiral structure \citep{Jorio08book}. In addition, the band gap of various  two-dimensional (2D) materials can also be tuned by the number of layers of the 2D materials \citep{Fantini04PRL,Neto09RMP,Wang12NN,Tran14PRB,Eperon14EES}. Although the range and variability of the band gaps of these (quasi) 1D and 2D systems are huge, as are their possible applications, their band gap is not continuously tuneable within a very wide range, and it varies from direct to indirect band gap by increasing the number of layers. Hence, a material with a tuneable direct band gap is highly desired, for which LCCs with alternating single and triple bonds are a perfect candidate \citep{Kertesz78JCP}.

The band gap for infinite LCCs (carbyne) in vacuum is calculated, and a wide variety of values ranging from 0.2 to 8.5 eV have been reported \citep{Yang06JPCA,Mostaani16PCCP}. Intrinsically, the band gap of an LCC depends on the BLA. Therefore, the large variety of predicted band gaps for carbyne can be explained by the difficulty of predicting the BLA of polyynes with density functionals that need to take into account both the electron-phonon coupling and many-electron interactions. The BLA decreases along with the increasing length of the LCC, which changes the electronic structure and results in a smaller band gap. However, this modulation by size has a fundamental limitation because a vanishing BLA can never be reached due to the Peierls distortion \citep{Peierls1955}, which means that the single-triple bonds can never be converted into double-double bonds. Thus, a finite band gap due to the saturation is expected for the carbyne.

Experimentally, free-standing LCCs need to be end--capped with hydrogen, adamantyl, trityl, tri--isopropylsilyl, or any of many other chemical groups to stabilize them. The longest end-capped LCCs synthesized so far consisted of 44 carbon atoms, with the band gap ranging from 4.7--2.6 eV corresponding to lengths of 4--44 carbon atoms respectively \citep{Chalifoux10NC,Agarwal13JRS,Eisler05JACS}.
For LCCs synthesized inside single--walled carbon nanotubes (SWCNTs), the resonance energies of LCCs with different lengths were reported to be 2.6--2.0 eV and were assigned to the dipole-forbidden transitions that become active via symmetry breaking by the CNT encapsulation \citep{Fantini06PRB,Malard07PRB,Moura09PRB,Andrade15C,Kang16C}.

Although the band gap of short LCCs has been well studied, the gap of LLCCs towards carbyne remains elusive as they are extremely unstable. In this work, we present the first experimental measurements of the band gaps of LLCCs stabilized within DWCNTs using resonant Raman excitation spectroscopy. Band gaps in the range of 2.253-1.848 eV have been observed. In addition, the feasibility of using long chains has allowed us to determine the smallest band gap of confined carbyne reported so far.

\section{\label{sec:level1}Experiments and Methods}

\begin{figure}[b]
\includegraphics[width=0.5\linewidth]{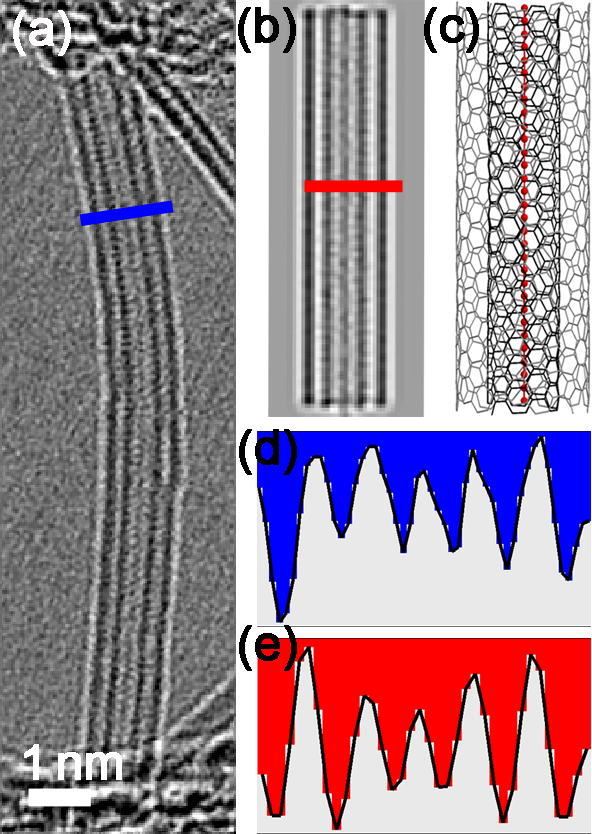}
\caption{\label{fig:tem} (color online). An experimental HRTEM image (a), a simulated HRTEM image (b), and a molecular model (c) of a LLCC@DWCNT. The line profiles for the experimental (d) and the simulated (e) LLCC@DWCNT at the corresponding marked positions are shown in blue (d) and red (e).}
\end{figure}

The LLCCs used in this study were synthesized inside DWCNTs with narrow inner diameters with a length ranging from about 30 carbon atoms (observed by transmission electron microscopy) up to more than 6000 carbon atoms (confirmed from near-field Raman spectroscopy) as described previously \citep{Shi16NM}. High-resolution transmission electron microscopy (HRTEM) was performed on a JEOL 2010F microscope conducted at 120 kV to avoid the LLCC decomposition. As shown in Fig.~\ref{fig:tem}, the HRTEM image and the corresponding simulations clearly confirm the hybrid structure of LLCC@DWCNT as the line profile consists of five peaks corresponding to two walls of a DWCNT and a LLCC in the middle. This unambiguously proves that the middle line represents a real LLCC and not a ghost contrast \citep{Hayashi06C}.
For this hybrid system with a LLCC longer than 10 nm (i.e. more than 80 carbon atoms) the distance between the tubes is almost the same as that to the LLCC, suggesting similar interactions. These interactions have to be considered for proper analysis of the following Raman spectra of LLCC@DWCNTs and the band gaps of LLCCs \citep{Wanko16PRB}.

Unfortunately, the band gaps for LLCCs@DWCNTs can not be investigated directly by absorption spectroscopy, as was done for the end--capped short LCCs, because the weak signal from the LLCCs is completely overlapped and covered by strong CNT absorption peaks \citep{Rohringer16AFM}. Therefore,  we applied resonant Raman excitation spectroscopy to obtain the lowest singlet excitation energy (i.e., the optical band gap) of LLCC@DWCNT. The experiments were performed under ambient conditions using triple monochromator Raman spectrometers (Dilor XY in Vienna, Dilor XY800 in Antwerp, and Horiba T64000 in Berlin) in combination with several tunable laser systems. Dye lasers with Rhodamine 110, Rhodamine 6G, and DCM are used to tune the laser wavelength from 540-570 nm, 560-620 nm, and 620-680 nm, respectively. A Ti:sapphire laser was also used to get the laser wavelengths between 680 and 770 nm.

Calculations were performed for H--terminated polyynes with 12--102 carbon atoms obtained using the second--order approximate coupled-cluster (CC2) method. This method is a size--consistent coupled--cluster approximation that avoids delocalization problems of density functional theory in extended systems. Our calculations are based on geometries optimized with SCS-MP2, which yields a similar BLA as the highly accurate CCSD(T) method \citep{Wanko16PRB}. The cc-pVDZ basis set and the turbomole software were used. 

\section{\label{sec:level1}Results and discussion}
\begin{figure*}[t]
\includegraphics[width=0.8\linewidth]{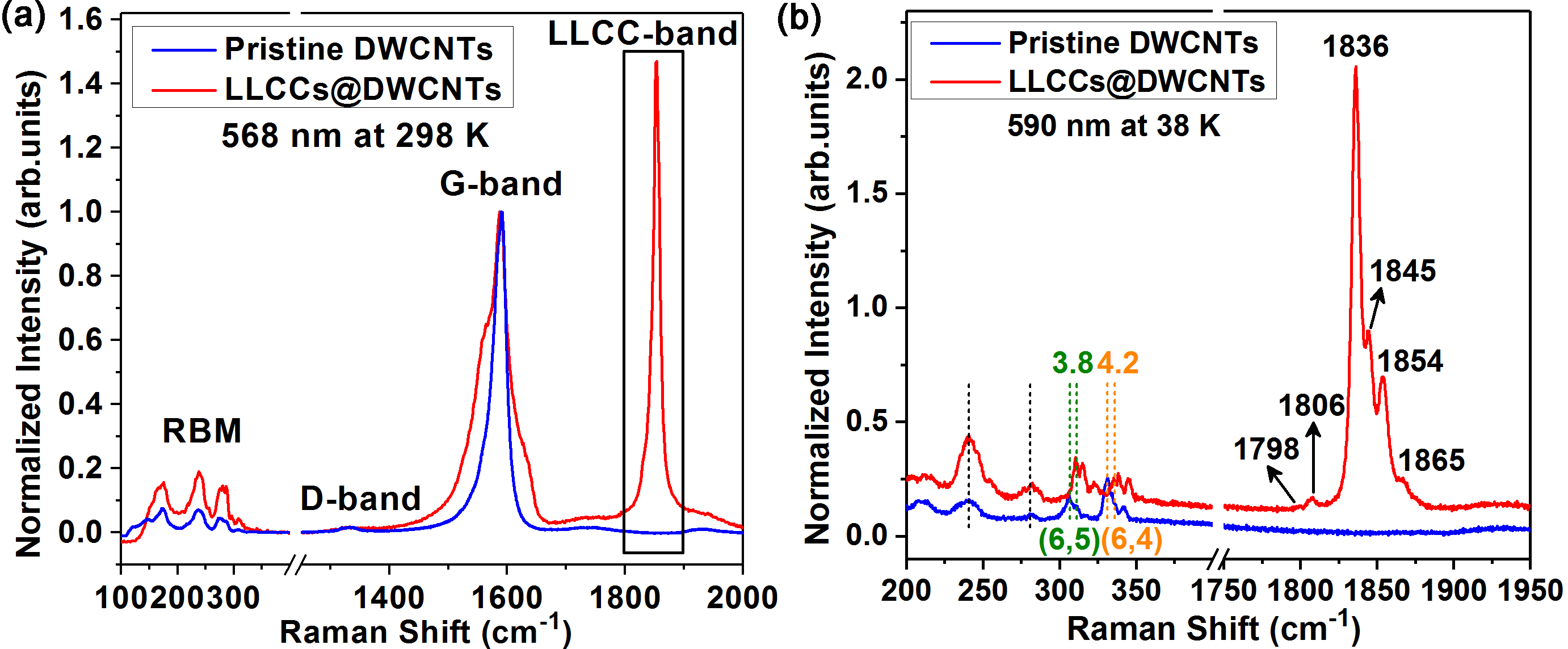}
\caption{\label{fig:Raman}(color online). (a) Raman spectra of pristine DWCNTs (blue line) and LLCC@DWCNTs (red line) measured with a 568 nm laser excitation at room temperature. The LLCC-band region is highlighted by a box. (b) Raman spectra of the same samples conducted with a 590 nm laser at 38 K. The colored dashed lines indicate the blue shift of 4.2 and 3.8 cm$^{-1}$ for the RBM peaks of (6,4) and (6,5) tubes, respectively, after encapsulation of the LLCCs. The black numbers are the frequencies of all observed Raman bands of the LLCCs.}
\end{figure*}

A typical Raman spectrum of LLCCs@DWCNTs consists of the Raman responses from the DWCNTs and the encapsulated LLCCs in the region at around 1850 cm$^{-1}$ (shown for 568 nm laser excitation in Fig.~\ref{fig:Raman}a). The line shape analysis of the LLCC-band measured for different excitation wavelengths obtained at room temperature (Fig. 3) and at an excitation wavelength of 590 nm at low temperature (Fig. 2b) reveals that it consists of 6 clearly resolved Raman peaks. Each of these different LLCC Raman frequencies corresponds to a resonance at a different excitation wavelength, hence corresponding to LLCCs with different band gaps. The observation of these different band gaps and Raman frequencies can either be attributed to LLCCs with different lengths \citep{Wanko16PRB} and/or due to different environmental interactions \citep{Shi16NM,Wanko16PRB}. Indeed, Fig. 2b shows that the radial breathing modes (RBMs) of the (6,5) and (6,4) inner tubes of LLCC-filled DWCNTs are blue-shifted with respect to the RBMs of freestanding DWCNTs, and both chiralities show a clearly different blue shift, indicating a different steric interaction between the chains and the tubes (similar to observed for water-filling \citep{Cambre10PRL}). Such van der Waals interactions, and also other interactions such as charge transfer between the LLCCs and host CNTs \citep{Wanko16PRB},  influence the BLA of the encapsulated chains and hence will result in different observed Raman frequencies (and as we demonstrate in this work, different band gaps).

The fact that we observe a discrete set of Raman features can be explained by the fact that only a limited number of possible inner tube diameters are available and are suitable for LLCC synthesis (only in the smallest diameters the ultra-long chains are stable \citep{Shi16NM}). This number is larger than 6, but the diameter of these inner tubes is distributed non-uniformly, and the line-width of the LLCC peaks is about 2 to 3 times broader than the LLCC peak from an individual LLCC@DWCNT obtained in near-field Raman spectra \citep{Shi16NM,Heeg17A}. This brings us to conclude that there are a few components in each of the six peaks observed.

Following the general trend that the longer the LCC is, the smaller the BLA and the resulting energy gap becomes, resonant Raman scattering is the ideal technique to identify the energy gap of ultra-long LCCs encapsulated within the DWCNTs.  Fig. 3 shows the Raman spectra as a function of laser excitation wavelength for different excitation ranges. For each of these excitation wavelengths (more than 50), we extracted the Raman intensities of the six LLCC bands that are resolved. The resonance Raman excitation profiles obtained are presented in Fig. 4. Note that two of the LLCC bands correspond to Raman frequencies around 1800 cm$^{-1}$, as we found previously for chains with lengths of several thousands of carbon atoms in near-field Raman spectroscopy \citep{Shi16NM}, sufficiently long for the LCC properties to have converged to those of infinitely long LCCs (carbyne).

\begin{figure*}[t]
\includegraphics[width=1\linewidth]{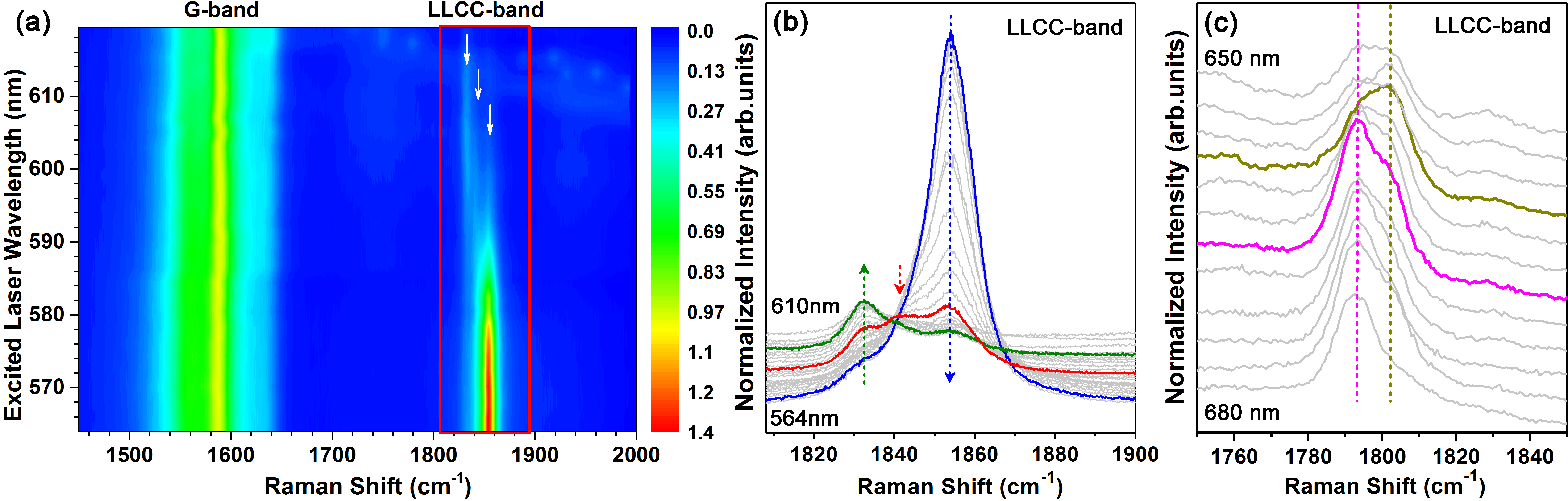}
\caption{\label{fig:spectra}(color online). (a) Resonance Raman mapping of LLCC@DWCNTs. The LLCC-band is highlighted by a red box. The three arrows point out three LLCC-band peaks with different frequencies. The evolution of the Raman spectra of LLCC@DWCNTs excited by lasers with the wavelength 564-610 nm (b) and 650-680 nm (c). The resonance Raman spectra for the LLCC peaks at 1793, 1802, 1832, 1842, as well as the sum of 1850 and 1856 cm$^{-1}$ were highlighted by magenta, dark yellow, olive, red, and blue lines, respectively.  }
\end{figure*}

\begin{figure}[t]
\includegraphics[width=1\linewidth]{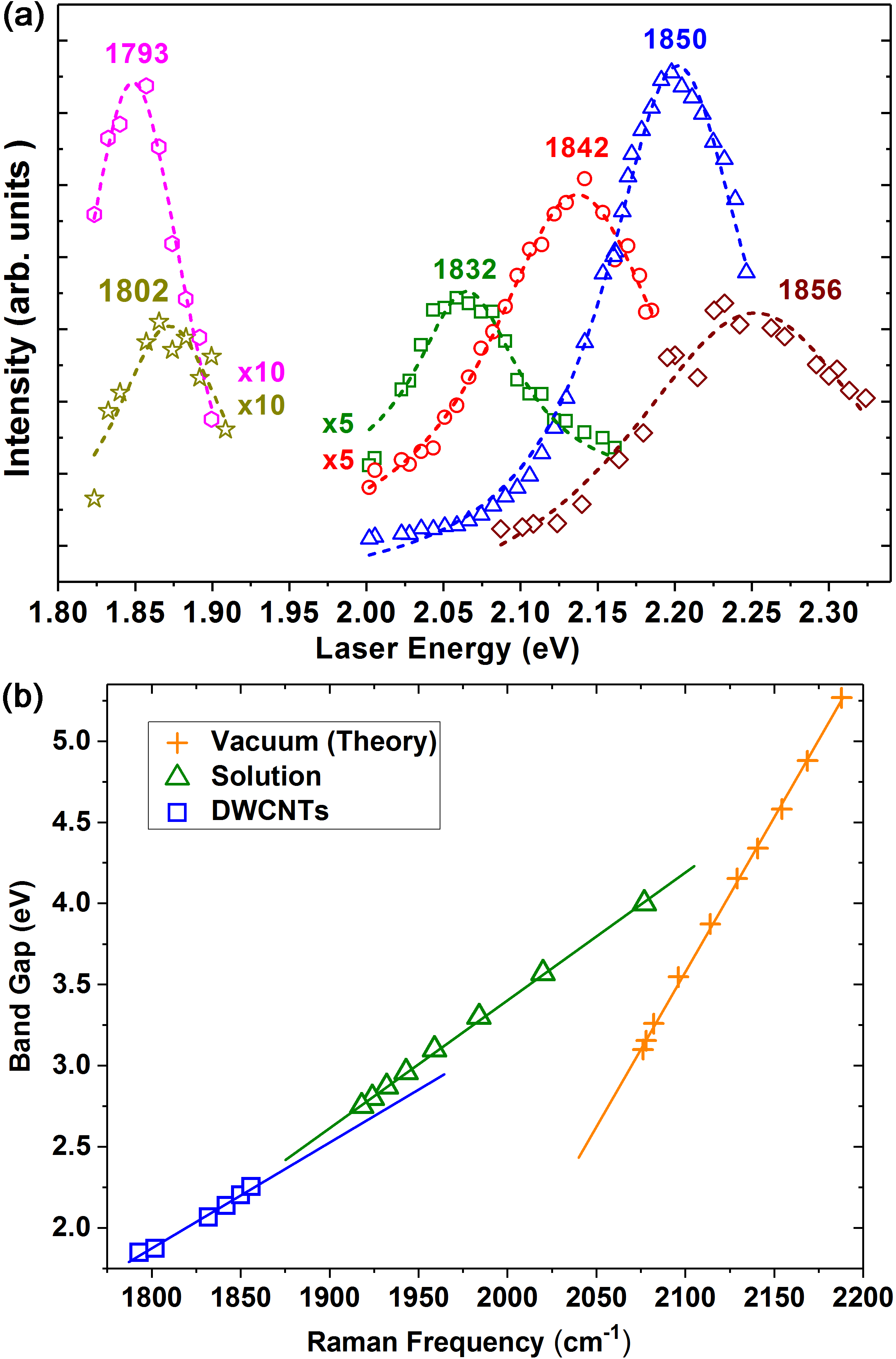}
\caption{\label{fig:profile} (color online). (a) Resonance Raman excitation profiles for six LLCC-band peaks. The dashed lines are a fit to the experimental data using Eq. (1). (b) Band gap of the LCCs as a function of  Raman frequency of the LCCs. The orange crosses are our theoretical prediction by ab-initio calculations on the free chains in vacuum, the olive triangles represent LCCs terminated by bulky end--groups in toluene (Raman frequencies) or hexane (band gap) \citep{Chalifoux10NC,Agarwal13JRS}, and the blue squares are our work on LLCCs inside DWCNTs. The linear lines are the fittings of the data points.}
\end{figure}

The analysis of the energy gaps from the resonance Raman excitation profiles was performed by fitting the separate Raman peaks in the LLCC-band and plotting their intensities as a function of excitation energy. The profiles for six LLCC Raman peaks are shown in Fig.~\ref{fig:profile}a with excitation energy steps smaller than 0.007 eV. These six profiles indicate strong electronic resonances that peak at different energy for each group of LLCCs, which matches the electronic energy gap of LLCCs. The profiles can be fitted by a semi-classical resonance Raman model \citep{Duque11AN,Haroz15PRB,Tran16PRB}:
\begin{equation}
I(E_{L}) \propto \left| {\frac{M}{E_{L}-E_{op}+i\frac{\Gamma}{2}}} \right|^{2}
\label{eq:raman-intensity}
\end{equation}
    where \textit{M, E$_{L}$, E$_{op}$, and $\Gamma$} are the incident resonance factor, laser excitation, optical transition, and an electronic broadening term, respectively. Note that a quantum model should be applied when considering both the incident and scattered resonances \citep{Peticolas70JCP}. This is not the case here, since we only measured the incident resonance. Applying Eq. (1) to fit the experimental profiles was done by adjusting \textit{M, E$_{op}$, and $\Gamma$}. The obtained \textit{E$_{op}$ and $\Gamma$} are summarized in Table 1. The widths of the excitation profiles are similar to the width of RBM or G-band excitation profiles in SWCNTs \citep{Duque11AN,Haroz15PRB,Tran16PRB}. The width of the peak located at 1856 cm$^{-1}$ is wider than the widths of the other peaks, indicating that the 1856 cm$^{-1}$ peak includes even more components than the others.

\begin{table}[b]
\caption{\label{tab:table1}%
Fitting analysis of resonance Raman excitation profiles.
}
\begin{ruledtabular}
\begin{tabular}{ccccccc}
Frequency(cm$^{-1}$) & 1793 & 1802 & 1832 & 1842 & 1850 & 1856\\
\colrule
\textit{E$_{op}$}(eV) & 1.848 & 1.872 & 2.065 & 2.137 & 2.202 & 2.253\\
\textit{$\Gamma$}(meV) & 72 & 84 & 116 & 131 & 97 & 145\\
\end{tabular}
\end{ruledtabular}
\end{table}

\textit{E$_{op}$} gives the band gaps. Figure 4b plots the band gap of the six resolved components as a function of Raman frequency (blue squares), and it compares this with previous experimental data of short LCCs obtained in solution and with our own theoretical calculations. All three data sets show a remarkably accurate linear dependence of the band gap and the Raman frequency, albeit with a different slope, demonstrating that the band gap is modulated in the same manner as the Raman frequency through the BLA \citep{Wanko16PRB}. Such linear relations between excitation energy, BLA, and vibrational frequencies are also known from other systems exposed to non-covalent interactions, e.g., the retinal chromophore inside different rhodopsin proteins \citep{Kochendoerfer97B,Fraehmcke12JPCB}. Note that for the solution data presented in Fig. 4b, optical band gaps were determined previously in hexane while Raman frequencies of the same chains were obtained in toluene, hence a different environment. The data for the encapsulated chains inside DWCNTs and the solution data were therefore fitted separately. 

\begin{figure}[t]
\includegraphics[width=1\linewidth]{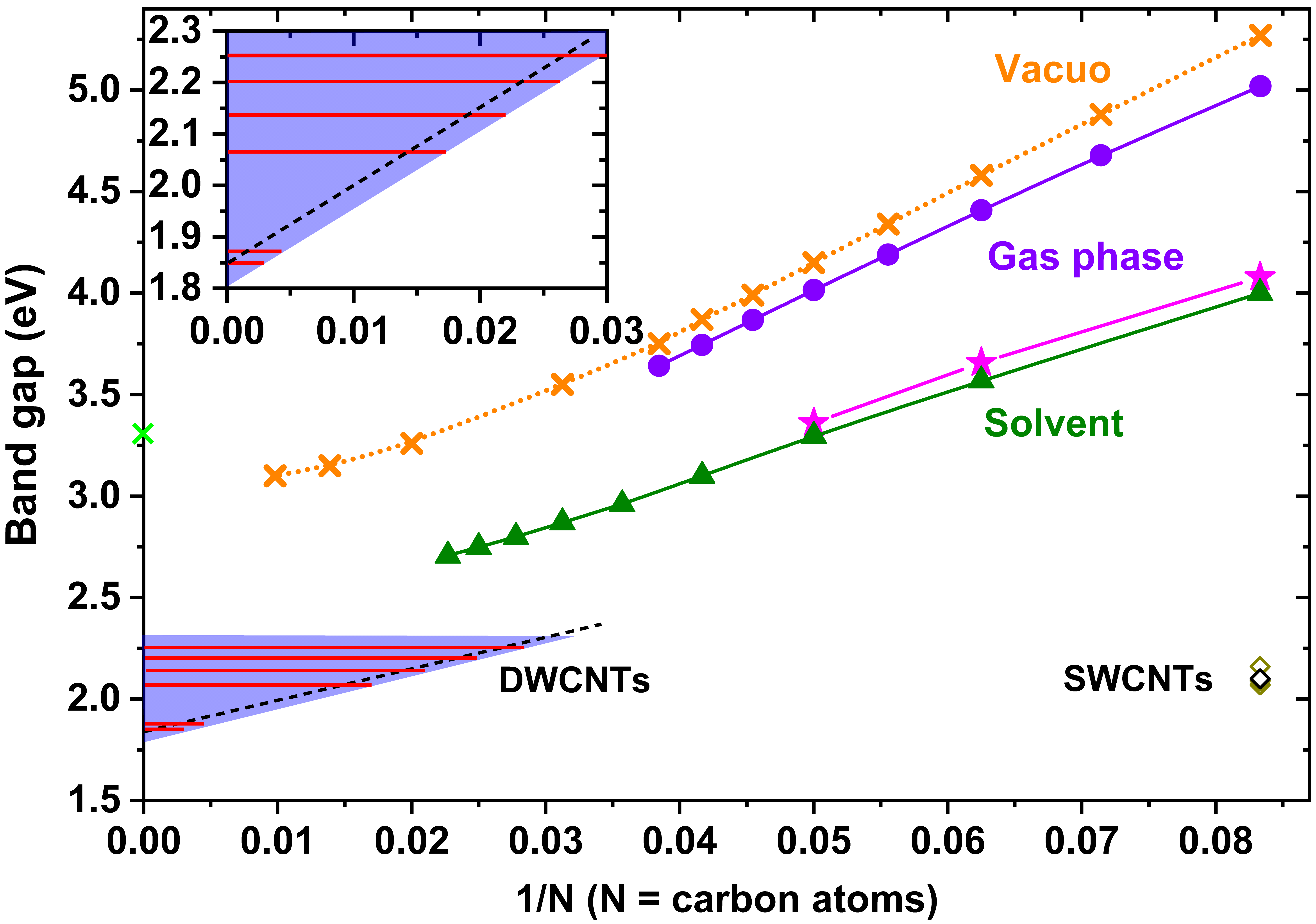}
\caption{\label{fig:energygap} (color online). The band gap as a function of inverse number of carbon atoms by ab-initio calculations (our work: orange crosses), predicted by Mostaani et al. for carbyne (green cross at 1/N=0) \citep{Mostaani16PCCP}, measured in gas phase (solid circles) \citep{Pino01JCP} or dissolved in a solvent (solid triangles and solid stars represent LCCs terminated by different chemical ending groups) \citep{Agarwal13JRS,Eisler05JACS} by absorption spectroscopy, and LCCs inside SWCNTs (open squares \citep{Nishide07JPCC,Moura09PRB}) / DWCNTs (our work: the light blue shadow) by resonance Raman spectroscopy. The data for SWCNTs does not represent band gap measurements but excited dark states \citep{Nishide07JPCC,Moura09PRB}. Inset: The enlarged part of the band gap of the confined chains. The dashed line is obtained by shifting the linear fit of the LCCs in solution, with the smallest observed band gap as the anchor point. The horizontal red lines represent the measured band gaps and the blue shaded area indicates the possible length range for our data.}
\end{figure}

It is remarkable that the accurate linear relation between the Raman frequency and the band gap holds over such a wide range, considering that many different factors are expected to lie at the origin of the variation of both. In general, the Raman frequency and the band gap are directly related to the BLA arising from the Peierls distortion of the LCC \citep{Peierls1955}. Indeed, previous previous theoretical and experimental work demonstrated that by applying strain to deliberately change the length of the carbon bonds, cumulene can be tuned from metallic (BLA = 0) to semiconducting (BLA $>$ 0), and finally it becomes insulating \citep{Liu13AN,Artyukhov14NL,LaTorre15NC}. Similarly, the band gap of graphene or other 2D materials can be adjusted by strain, type of stacking, charging the substrate, chemical functionalization, electronic doping, etc \citep{Neto09RMP,Wang12NN,Tran14PRB,Eperon14EES}. Previous studies have demonstrated that the BLA not only depends on the intrinsic length of the LCCs \citep{Yang06JPCA}, but also on extrinsic factors such as environment interactions (van der Waals interactions, charge transfer, and dielectric screening) \citep{Yang06JPCA,Wanko16PRB,Nishide07JPCC,Moura09PRB} and the specific choice of chemical groups at the end of the chains \citep{Agarwal13JRS,Eisler05JACS}. The relative contribution of each of these effects in our experimental data is however difficult to disentangle and will be different for different length ranges. For short chains, with lengths ranging from 6 to 44 carbon atoms, as previously measured in solution \citep{Agarwal13JRS,Eisler05JACS} or in the gas phase \citep{Pino01JCP}, it is well known that the length of the chains strongly influences the BLA and hence the Raman frequency \citep{Wanko16PRB}. Indeed, when plotting the experimentally-determined band gaps of those chains with well-defined lengths as a function of the inverse number of carbon atoms, a linear dependence on 1/N is obtained, with N the number of carbon atoms in the chain (solid triangles, squares, and circles in Fig. 5). To extend the band gap dependence on length to longer chain lengths than those measured in the gas phase, we also included our calculated excitation energies of H-terminated polyynes with 12-102 carbon atoms (orange crosses in Fig. 5). The excitation energies for the lowest allowed singlet transition, as presented in Fig.~\ref{fig:energygap}, agree very well with the gas--phase measurements \citep{Pino01JCP}, and they show a deviation from the linear dependence for the longest chains, approaching the band gap of 3.3 eV of carbyne obtained from the diffusion Monte Carlo result of Mostaani et al. \citep{Mostaani16PCCP}. Interestingly, the LCCs measured in solution show a significant downshift of the band gap energy with respect to the gas--phase measurements, indicating that the band gap is very sensitive to the environment through van der Waals interaction, dielectric screening, and/or a charge transfer interaction. In addition, the groups that are at the end of the chains (end-caps) also influence the BLA. As shown in Fig. 5 the solid triangles and stars represent chains that are terminated with different end--groups while surrounded by the same solvent, resulting in a shift of about 0.1 eV \citep{Agarwal13JRS,Eisler05JACS}. In particular for short chains one can expect a strong influence of these end--groups on the band gap of the chains \citep{Weimer05CP}, which can be used to tune this band gap to some extent \citep{Milani17JPCC}. Implementing our measured band gaps into Fig. 5 (red horizontal lines represent the measured band gaps) is not so straightforward, as the actual lengths of the LLCCs in our samples vary from 30 up to more than 6000 atoms \citep{Shi16NM}. From near-field Raman spectroscopy, it is known that at least for LLCCs with lengths longer than 30 nm (i.e. N > 230) \citep{Heeg17A} the interactions with the environment (i.e. the chirality of the surrounding inner CNT) dominate over length in determining the vibrational frequency.

Combining this information for short and long chains, we can define the upper and lower bounds to the lengths of the LLCCs in our experimental data. If, in one limiting case, we assume that all the Raman frequencies in our experiments originate from chains longer than 30 nm (1/N approaching 0), and hence the length hardly influences the Raman frequency, then the 6 different Raman peaks that are observed can be attributed to interactions with different inner tube chiralities. However, if we assume that also shorter chains contribute to the observed Raman frequencies, as these are also present in our samples \citep{Shi16NM}, then length can also contribute significantly to the observed band gap variations. To define the upper limit in $1/N$, one can assume all chains encapsulated in the same CNT environment, i.e., the one that yields the largest possible downshift. Then the observed band gaps would depend solely on the chain length. Hence we take a similar length dependence as obtained for the short chains in solution (i.e. similar slope), but we shift it to lower energy with the lowest measured band gap as the anchor point for the longest chains (dashed line in Fig. 5). Note that the shift to lower energy for chains encapsulated in DWCNTs can be explained in the same manner as the shift from gas phase to solution spectra. Most likely, in our samples including both shorter and ultra-long chains, chain lengths are in between these two limits (augmented with an experimental error margin), as highlighted by the shaded area in Fig. 5. If the exact curve saturates for large N, like our theoretical gas-phase results, the shaded area would extend further to the right, but the limiting values for the band gap of confined carbyne would be unaffected. The smallest band gap measured in our samples is 1.848 eV and is much lower than the value of the band gap of carbyne (2.56 eV), which was extrapolated from the band gap of short chains with lengths only ranging from 6 to 44 carbon atoms \citep{Chalifoux10NC,Agarwal13JRS}. However, our results demonstrate that such an extrapolation is difficult to perform, as for long chains the length is not longer a determining factor for the band gap, and in particular also the interaction with the environment needs to be taken into account.

Our experimental data cannot resolve which interaction dominates. The chirality-dependent blue-shift of the RBMs of LLCC-filled CNTs with respect to pristine CNTs (Fig. 2b) suggests a steric interaction between the LLCCs and the CNTs, similar as observed previously for water-filling \citep{Wenseleers07AM,Cambre10PRL}. Previous experimental studies also reported that the band gap of the LCC inside MWCNTs does not depend on the number of host CNT walls \citep{Zhao03PRL,Fantini06PRB,Shi11NR}, rather it depends on the diameter of the inner--most tubes. In addition, when encapsulating the same $C_{10}H_2$ chain inside three different SWCNT diameter distributions, an energy shift of the order of 0.1 eV was observed as a consequence of different dielectric screening \citep{Moura09PRB}. Indeed, a recent theoretical work showed that, apart from charge transfer \citep{Yang06JPCA}, van der Waals interactions strongly affect the electronic structure, BLA, and vibrational properties of encapsulated polyynes \citep{Wanko16PRB}. Overall, it is still a big challenge to quantitatively evaluate the effects of van der Waals interaction, dielectric screening, or charge transfer on the hybrid LCC@CNT system theoretically or experimentally.

\section{\label{sec:level1}Conclusions}

In summary, the band gaps of confined LLCCs were directly measured by resonance Raman excitation spectroscopy. An accurate linear relation between Raman frequency and band gap was obtained. The LLCCs inside DWCNTs possess band gaps of 2.253--1.848 eV. The band gap of 1.848 eV for the long confined LCCs is the smallest band gap observed so far. Note that the band gap values reported here are the optical band gaps, and thus include (reduction by) the exciton binding energy, as is the case also in all previous measurements on short chains. Theoretical calculations show the exciton binding energy of carbon chains is rather small (about 0.1 eV) \citep{Mostaani16PCCP}. Our results illustrate the theoretical challenges of taking into account the interactions with the environment to calculate the band gap of LCCs. LCCs were predicted to be the stiffest materials \citep{Liu13AN}, and they can even be used for spin transport \citep{Zanolli10AN}. Also, the LCC@CNT system can achieve metallic transport properties by a high density of states at the Fermi level, due to a combined effect of orbital hybridization and charge transfer \citep{Rusznyak05PRB,Tapia10C}. Together with the tunable band gap, LCCs would be a promising candidate for future nanoelectronic, photonic, and spintronic devices.

\begin{acknowledgments}
  This work was supported by the Austrian Science Funds (FWF, P27769-N20) and the EU project (2D-Ink FA726006). L.S. thanks the scholarship supported by the China Scholarship Council. A.R. acknowledges financial support from the European Research Council (ERC-2015-AdG-694097), Grupos Consolidados (IT578-13), and NOMAD (GA no.676580 ). P.R., S.C., and W.W. acknowledge funding from the Fund for Scientific Research Flanders, Belgium (FWO, projects No. G040011N, G021112N, 1513513N and 1512716N), which also supported S.C. through a postdoctoral fellowship. S.C. also acknowledges funding from European Research Council Starting Grant No. ERC-2015-StG-679841. We thank Kazu Suenaga and Yoshiko Niimi for the HRTEM measurements, Hans Kuzmany for insightful discussion, and Patryk Kusch for Ti:sapphire laser preparation.
\end{acknowledgments}


\end{document}